\documentclass[10pt]{iopart}
\usepackage{iopams}

\usepackage{graphicx}
\usepackage{epsfig}		
\usepackage{dcolumn}
\usepackage{bm}
\begin{document}

\title[A simple kinetic approximation for heavy meson mass spectroscopy]
{The charmonium and bottomonium mass spectroscopy
with a simple approximation of the kinetic term}

\author{A. E. Bernardini \footnote[1]{E-mail address: alexeb@ifi.unicamp.br} and C. Dobrigkeit \footnote[2]{E-mail address: carola@ifi.unicamp.br}  
}

\address{

Department of Cosmic Rays and Chronology, State University of Campinas,
PO Box 6165, 13083-970, Campinas, SP, Brazil
}

\begin{abstract}
In this paper we propose a particular description of meson spectroscopy, with emphasis in heavy bound states like charmonia and bottomonia, after working on the main aspects of the construction of an effective potential model. 
We use the prerogatives from ``soft QCD'' to determine the effective potential terms, establishing the asymptotic Coulomb term from one gluon exchange approximation. 
At the same time, a linear confinement term is introduced in agreement with QCD and phenomenological prescription.
The main aspect of this work is the simplification in the calculation, consequence of a precise and simplified description of the kinetic term of the Hamiltonian.  
With this proposition we perform the calculations of mass spectroscopy for charmonium and bottomonium mesons and we discuss the real physical possibilities of developing a generalized potential model, its possible advantages relative to experimental parameterization and complexity in numerical calculations.
\end{abstract}


\submitto{\jpg}

\maketitle

\section{\label{sec1}Introduction}

Since the discovery of QCD (Quantum Chromodynamics), there have been remarkable technical achievements in perturbative calculations applied to hadrons. 
However, it is difficult to use QCD directly to compute hadronic properties. 
In this context, phenomenological potential models have provided extremely satisfactory results in describing ordinary hadrons, more specifically quark - antiquark bound states (mesons). 

The pioneer work proposing a potential model inspired in QCD was the De R\'{u}jula, Georgi and Glashow model \cite{ruj}.
Since then, the phenomenological prerogatives to characterize potential models were based on:
($i$) hadrons are in a color singlet;
($ii$) at short distances, QCD interactions obey asymptotic freedom and one gluon exchange is dominant with a limited coupling constant (potential will be of Coulomb type);
($iii$) at long distances, the quarks are bound by a flavor independent long range potential, obtained phenomenologically;
($iv$) quarks in a bound state move relativistically and the approximation to first order in $v^2 \backslash c^2$ is normally used.

Among several effective models worked out until now \cite{ruj, bet, sal, gro, qui, god, luc, ban, sem, pau, bra}, the less complex alternative involves the Schr\"{o}dinger formalism, that can also be derived from the Salpeter \cite{sal} description of a bound state (\ref{eq1}):
\begin{eqnarray}
\fl
[E-H_1-H_2]\tilde{\phi}(\bi{p},P)= & \int d^3 p'  \{ \Lambda_1^+ \gamma_1^0 U(\bi{p}-\bi{p'})\tilde{\phi}(\bi{p'},P)\gamma_2^0 \Lambda_2^+ \nonumber\\ 
& ~~~~~~~~~~~~~~~~~~~~~~ -  \Lambda_1^- \gamma_1^0 U(\bi{p}-\bi{p'})\tilde{\phi}(\bi{p'},P)\gamma_2^0 \Lambda_2^- \},
\label{eq1} 
\end{eqnarray}
where
\begin{equation}
{\Lambda _i}^{\pm}=\frac{1}{2}\left({1 \pm \frac{H_i}{\sqrt{{\bi{p_i}}^2+{m_i}^2}}}\right),
\label{eq2}
\end{equation}
\begin{equation}
\tilde{\phi}(\bi{p},P)=\int{d p^0}\tilde{\psi}(p^0,\bi{p},P),
\label{eq3}
\end{equation}
and $H_i$ is the Dirac Hamiltonian.

From $H_i$ we obtain the kinetic approximation term $T_i^{kin}$ and with the following approximations (\ref{eq4}) the perturbative form of Schr\"{o}dinger equation (\ref{eq5}) is obtained:
\begin{equation}
\label{eq4}
\cases{
\Lambda_i^+ U(\bi{p}-\bi{p'})\tilde{\phi}(\bi{p'})\Lambda_i^+ \longrightarrow & $U(\bi{p}-\bi{p}'){\phi}(\bi{p'})$\\ 
\Lambda_i^- U(\bi{p}-\bi{p'})\tilde{\phi}(\bi{p'})\Lambda_i^- \longrightarrow & $0$\\},
\end{equation}
\begin{equation} 
\left[\sum_{i=1}^2T_i^{kin} - E \right]\tilde{\phi}(\bi{p'})=-\frac{1}{(2\pi)^3}\int{d^3 \bi{p'}\,\tilde{U}(\bi{p}-\bi{p'})\tilde{\phi}(\bi{p'})}. 
\label{eq5} 
\end{equation} 

The physical prerogatives impose an effective Hamiltonian only with vector and scalar Lorentz components.
It includes all Breit - Fermi interaction terms \cite{gro, bbr, fer}, consequently, all spin effect terms.

\section{\label{sec2}The Hamiltonian kinetic term} 
 
Since the pioneer papers of Stanley and Robson \cite{sta} and Godfrey and Isgur \cite{god} the kinetic term effects of a quark - antiquark bound state are observed with emphasis to relativistic effects, although treating heavy quark systems. 
Keeping in mind the perturbative treatment of relativistic corrections, despite its effects making the wave functions result more complicated \cite{qui, ld1, ful, ld2}, we determine a kinetic energy term approximation that describes the mean behavior for low energy states.
We adopt a nonrelativistic approximation based on Martin inequality (\ref{eq6}) as suggested in \cite{ld2, mar}: 
\begin{equation} 
\sqrt{\bi{p}^2 +m^2} \leq \frac{M}{2}+\frac{\bi{p}^2}{2M}+ \frac{m^2}{2M}. 
\label{eq6} 
\end{equation} 
 
For a given $M$, the equality in (\ref{eq6}) is obtained at the point $\bi{p}_0=\sqrt{M^2 - m^2}$ in the momentum space.
The curves defined by the two sides of equation (\ref{eq6}) are tangent at $\bi{p}_0$, suggesting $M^2=m^2+{\bi{p}_0}^2$.

Since Martin's condition is an operator relation, that inequality will continue valid for mean values of a state and for the average of mean values over a set of states with arbitrary $M$ and $\bi{p}_0$.
In \cite{ld2} the choice of ${\bi{p}_0}^2=$ $<\bi{p}^2>$ was suggested by Jaczko and Duran for a set of states, with a reasonably exact nonrelativistic approximation.
The equation (\ref{eq6}) can be rewritten in the following way:
\begin{equation} 
\sqrt{\bi{p}^2 +m^2} \leq M + \frac{\bi{p}^2}{2M}- \frac{<\bi{p}^2>}{2M}, 
\label{eq7}
\end{equation}
with
\begin{equation} 
M=\sqrt{<\bi{p}^2> +m^2}.
\label{eq8} 
\end{equation}
 
The physical content of this result can be illustrated through the following direct expansion around ${\bi{p}_0}^2$ (\ref{eq9}):
\begin{eqnarray}
\sqrt{\bi{p}^2 +m^2} &=& \sqrt{\bi{p}^2-{\bi{p}_0}^2 +M^2} \nonumber\\
& = & M + \frac{\bi{p}^2-{\bi{p}_0}^2}{2M}- \frac{\left(\bi{p}^2-{\bi{p}_0}^2\right)^2}{8M^3}+...\,. 
\label{eq9} 
\end{eqnarray} 

We observe that an approximately correct choice of the parameter ${\bi{p}_0}^2$ will provide an excellent result for mass spectroscopy of heavy quarkonium systems.
In \cite{ld2} a linear approximation in $\bi{p}^2$ with a translational term $\epsilon$ was adopted and the parameter ${\bi{p}_0}^2$ is suggested and calculated to be the average of the mean square values of momenta over the first four energy states.

The kinetic energy of each particle in the bound state:
\begin{equation} 
{T^{kin}}_{(exact)}=\sqrt{{\bi{p}}^2 +{m}^2} 
\label{eq10} 
\end{equation}
can be approximated by an expression which is linear in ${\bi{p}}^2$ as: 
\begin{equation} 
{T_i^{kin}}_{(app.)}=\sqrt{{\bi{p}_{0}}^2 +{m}^2} +\frac{{\bi{p}}^2-{\bi{p}_{0}}^2}{2\sqrt{{\bi{p}_{0}}^2 +{m}^2}}. 
\label{eq11} 
\end{equation}

Although this parameterization has the same structure as that adopted by Jaczko and Duran in \cite{ld2}, we followed a different procedure to calculate the values of ${\bi{p}_0}^2$.
The values of ${\bi{p}_0}^2$ are determined as to minimize the mean square deviation (MSD) of the linear approximation (\ref{eq11}) from the exact relativistic kinetic term (\ref{eq10}).
We can visualize the results of the MSD procedure in figures 1.(a) and 1.(b), in the interval $0 \, \mbox{GeV}^2 \, < \, \bi{p}^2 \, < \, 10 \, \mbox{GeV}^2$, depending on ${\bi{p}_0}^2$ for different values of mass interesting in the study of heavy quark systems (from $0 \, \mbox{GeV}$ to $5 \,\mbox{GeV}$).
\begin{figure}[h]
\begin{center}
\epsfig{file= 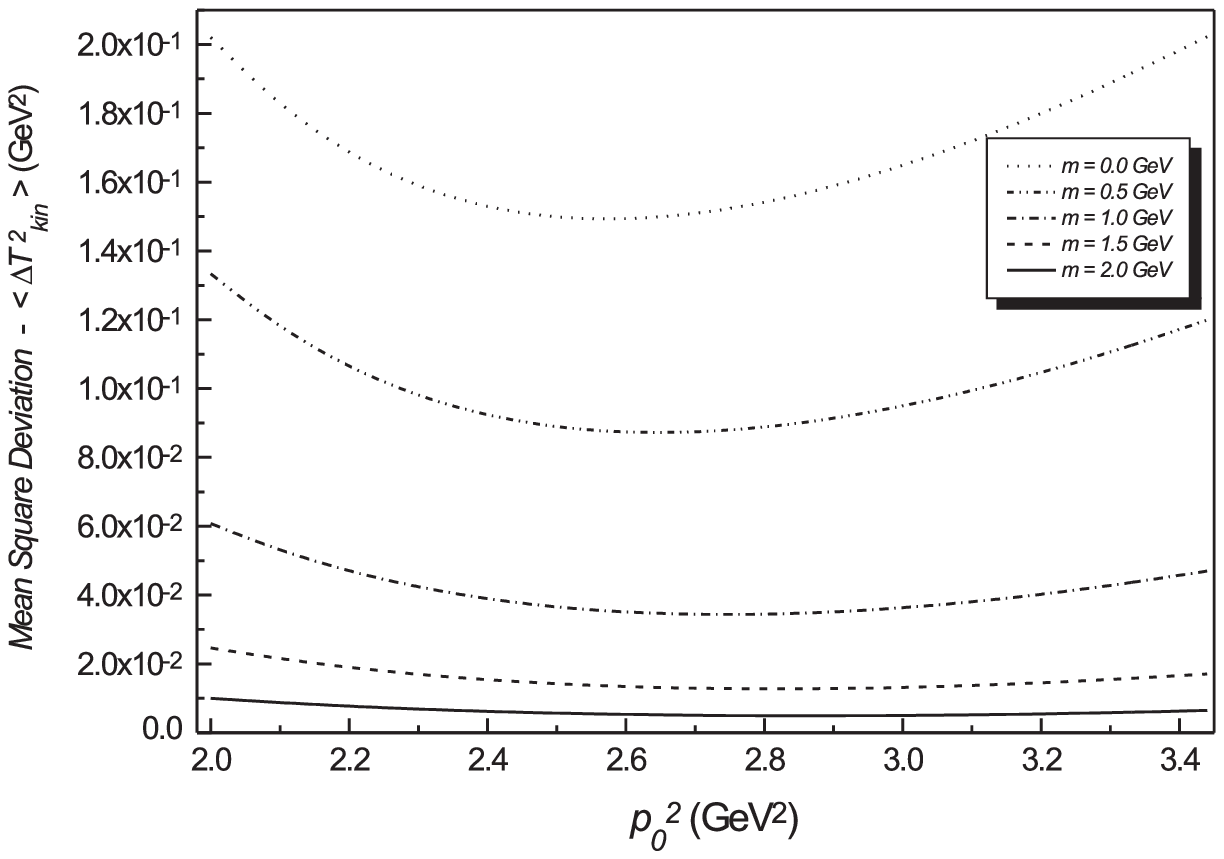,  height=8.6 cm, width=9 cm}
\label{fig1}
\caption{\textbf{(a)} Mean square deviation of exact and approximated kinetic term for masses between $0$ and $2.0 \, \mbox{GeV}$.}
\epsfig{file= 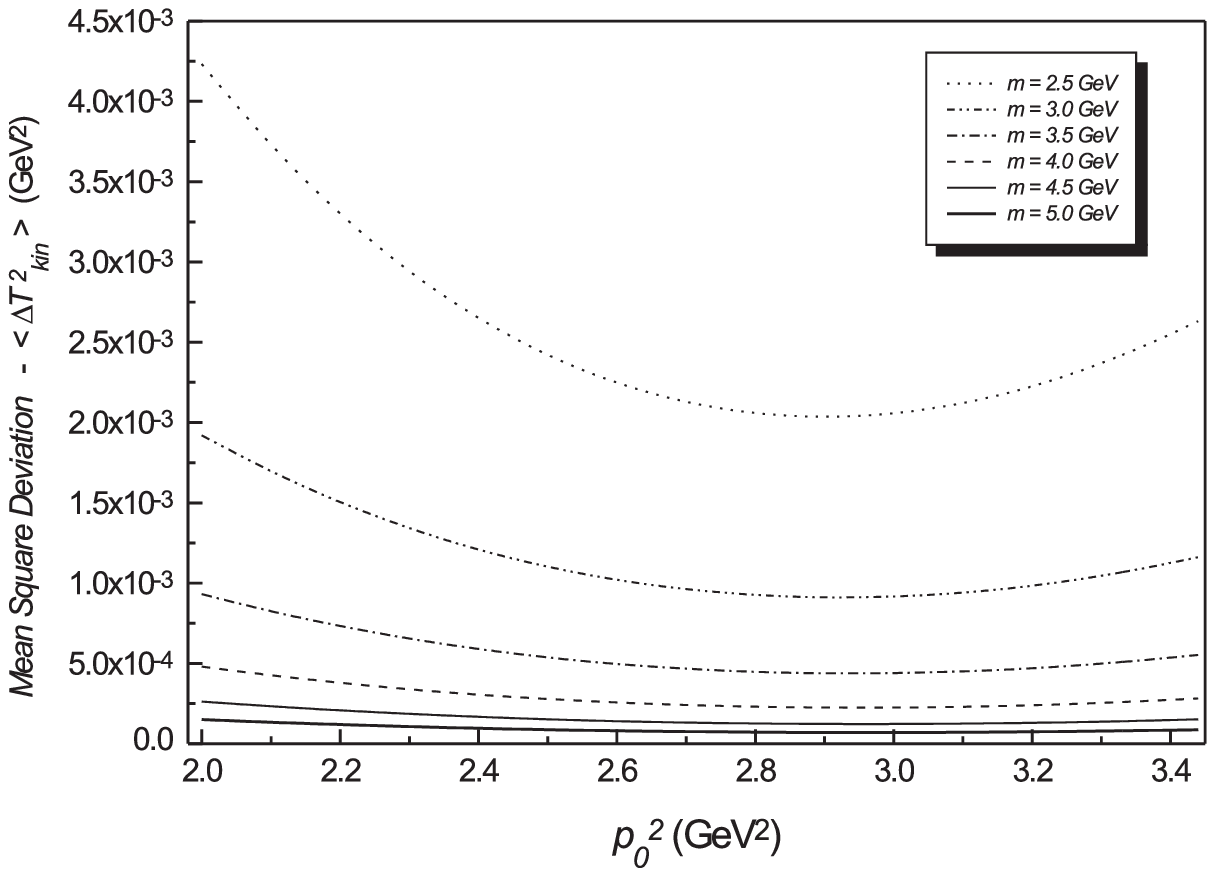,  height=8.6 cm, width=9 cm} 
\setcounter{figure}{0}
\label{fig1b}
\caption{\textbf{(b)} Mean square deviation of exact and approximated kinetic term for masses between $2.5$ and $5 \, \mbox{GeV}$.}
\end{center}
\end{figure} 
\begin{figure}
\begin{center}
\epsfig{file= 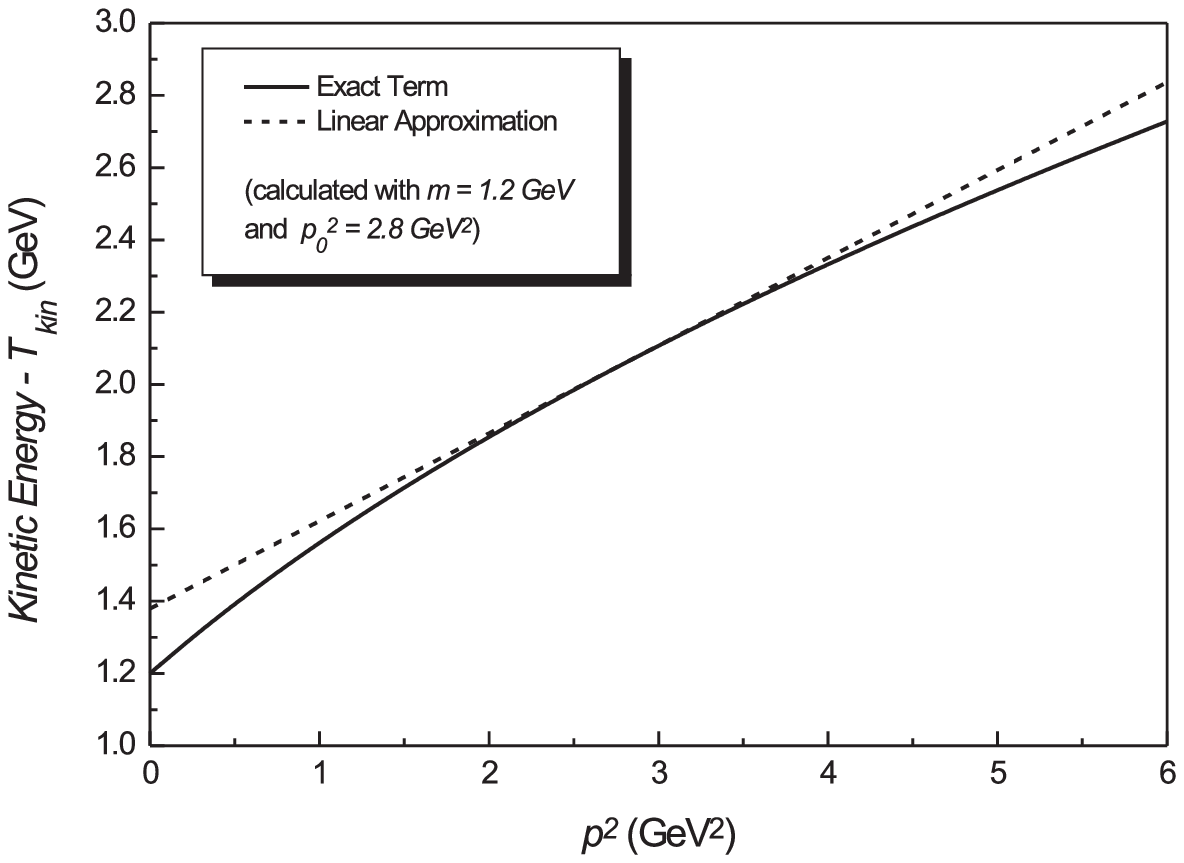,  height=9 cm, width=9 cm}
\label{fig2}
\caption{\textbf{(a)} Linear approximation and exact term \textit{versus} $\bi{p}^2$ with $\bi{p}_0^2 = 2.8 \, \mbox{GeV}^2$, for $m = 1.2 \, \mbox{GeV}$.} 
\epsfig{file= 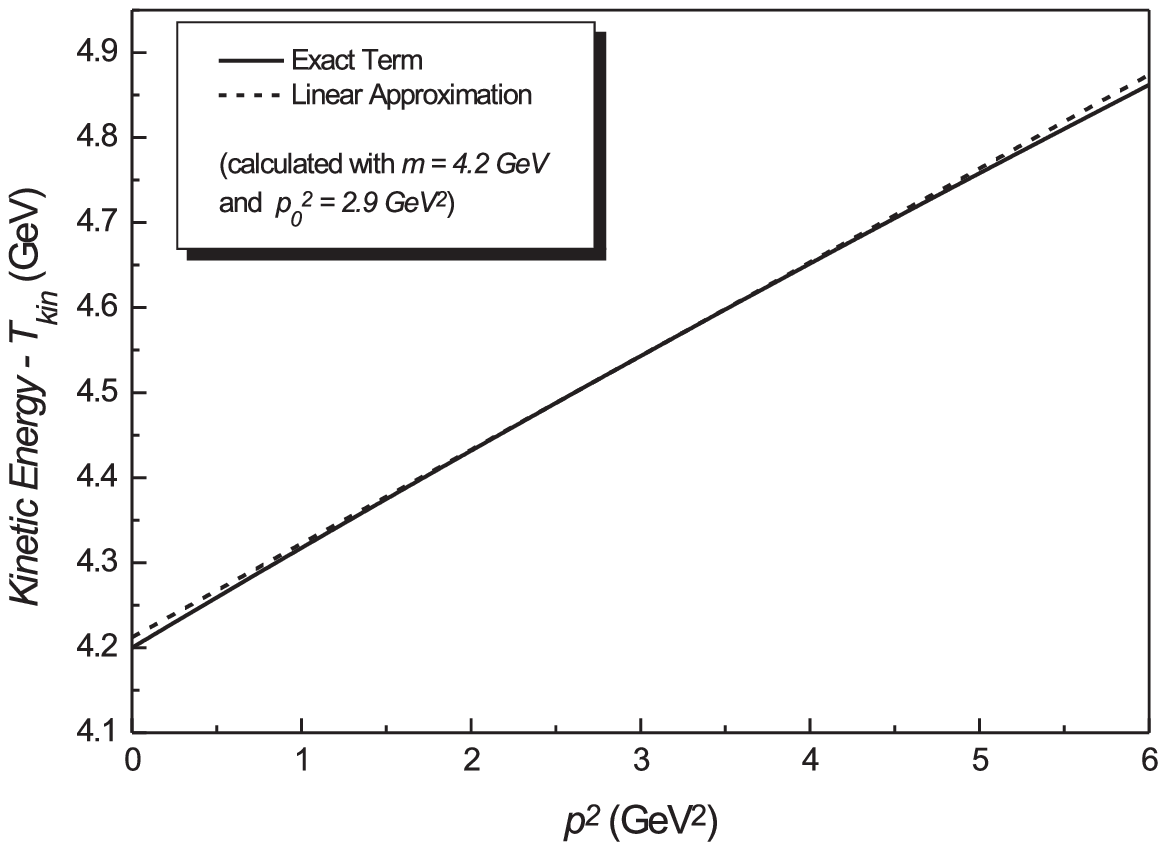,  height=9 cm, width=9 cm}
\setcounter{figure}{1}
\label{fig2b} 
\caption{\textbf{(b)} Linear approximation and exact term \textit{versus} $\bi{p}^2$ with $\bi{p}_0^2 = 2.9 \, \mbox{GeV}^2$, for $m = 4.2 \, \mbox{GeV}$.} 
\end{center} 
\end{figure}

The advantage of such procedure is that the obtained values of ${\bi{p}_0}^2$ are independent from any energy state used as input data.
Consequently, they can be applied to calculating the whole mass spectrum of a bound state while in \cite{ld2} the calculation is not extended beyond the first four energy states of a bound state.

Expression (\ref{eq11}) will be extremely accurate in the region of kinetic energy where the probability of finding a quark in a bound state is significant.
This region corresponds to an intrinsic energy of each particle $\bi{p}^2 \, < \, 10 \, \mbox{GeV}^2$ in the center of momentum frame.

We note that the value of the parameter ${\bi{p}_0}^2$ is approximately $2.9\, \mbox{GeV}^2$ for a particle with mass between $4.0 \, \mbox{GeV}$ and $5.0 \, \mbox{GeV}$, as is the case of the bottom quark.
In the case of the charm quark, with mass between $1.0 \, \mbox{GeV}$ and $2.0 \, \mbox{GeV}$, the value of the parameter ${\bi{p}_0}^2$ is approximately $2.8\, \mbox{GeV}^2$. 

Figures 2.(a) and 2.(b) show the exact kinetic term and its linear approximation for these two cases, using typical values of masses in the regions mentioned above and for the main region of intrinsic momenta.

The procedure adopted here avoids sistematic errors induced by simultaneous evoluation of the parameter ${\bi{p}_0}^2$ and the effective quark masses.

Other treatments of the kinetic nonlocal terms have been already proposed \cite{god, bra, ful, ld2, ld3}, however, none of them provides so simple subsequent calculations.   
 
\section{\label{sec3}The interacting potential} 
 
Our propositions will follow some prescriptions that were already developed in \cite{god, luc}.
We start with the Coulomb plus Linear form of the interacting potential without perturbative corrections.

We know that QCD provides expression (\ref{eq12}) for the Coulomb coupling constant through a first order approximation following Feynman diagrams \cite{web}:
\begin{equation} 
\alpha _s(k^2)=\frac{12 \pi}{\left(33-2n_f\right)\ln\left(\frac{k^2}{\Lambda^2}\right)}. 
\label{eq12} 
\end{equation} 
 
The scale parameter $\Lambda$ lies in the interval between  $0.15 \, \mbox{GeV}$ and $0.30 \, \mbox{GeV}$ when the flavor number ($n_f$) is  4 or 5 \cite{web}.
Observing the behavior of $\alpha _s(k^2)$, it approximates to a constant value in the region of $k$ where $5 \, \mbox{GeV} < k < 20 \, \mbox{GeV}$.
However, when $k\longrightarrow\Lambda$, the first order perturbative treatment is not possible since (\ref{eq12}) diverges indicating the confinement. 

Godfrey and Isgur \cite{god} proposed a parameterization of such behavior for $5 \, \mbox{GeV} < k < 20 \, \mbox{GeV}$ in the following way:
\begin{equation} 
\fl
\alpha _s(k^2)=\sum _m \alpha _m \exp{\left(-\frac{k^2}{\eta _m^2}\right)}~~\Rightarrow~~ 
\alpha _s(r)=\frac{2}{\sqrt{\pi}}\sum _m \alpha _m \int _0^{\frac{\mbox{{\small w}}_m \mbox{{\small $r$}}}{2}} \exp{\left(-x^2\right)} dx. 
\label{eq13}
\end{equation}

A simpler approximation can be suggested in the treatment of heavy quark systems:
\begin{equation} 
\alpha _s(r)\approx \alpha _s \left[1- \exp\left(-\mbox{w} r\right)\right]. 
\label{eq15} 
\end{equation} 

This parameterization correctly describes the asymptotic behavior of the coupling constant.
Consequently we may assume it is correct just for the asymptotic region where the perturbative treatment may be applicable.

\begin{figure}
\begin{center} 
\epsfig{file= 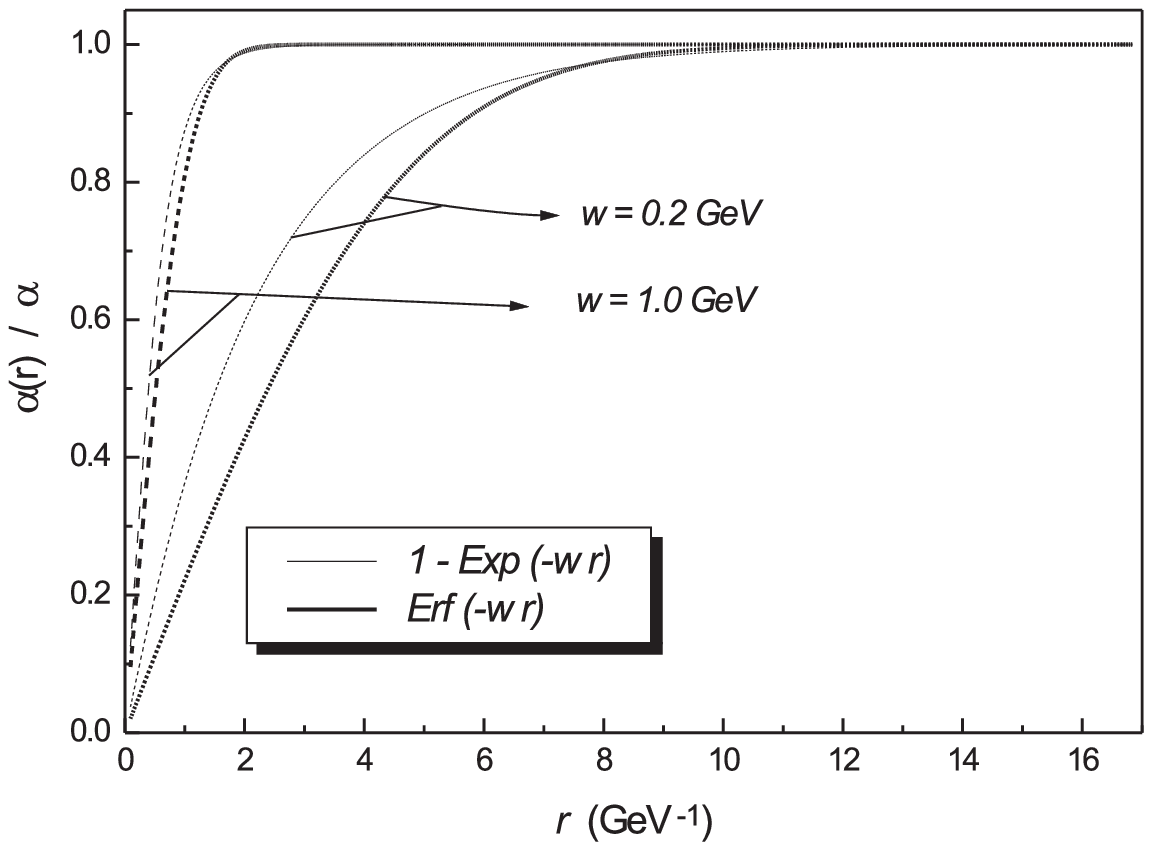,  height=9 cm, width=9 cm} 
\caption{\label{fig3}Coulomb coupling constant parameterization in the use of approximation (\ref{eq15}).} 
\end{center}
\end{figure}

The confining term will turn up in agreement with experimental data \cite{luc} that indicate a linear behavior.
In the interaction region $5 \, \mbox{GeV} < k < 20 \, \mbox{GeV}$ just the first term of the summation (\ref{eq13}) is sufficient to describe the interaction potential (\ref{eq16}), and in function of the parameter $\mbox{w}$ we can reproduce this behavior through the equation (\ref{eq15}).
This procedure is illustrated in figure \ref{fig3}.
\begin{equation} 
U(r)=-\frac{4}{3}\frac{\alpha_s}{r}\left[1- \exp\left(-\mbox{w} r\right)\right] + a r + b. 
\label{eq16} 
\end{equation}

Obviously, expression (\ref{eq16}) just describes the pure terms of the potential $(U(r)\,=\,\mbox{Scalar(r)}\,+\,\mbox{Vector(r)})$, but in our calculation all interacting elements with orbital angular momentum, spin and nonlocal term dependencies were comprised following the Gromes prescription \cite{gro, luc, muk} for the study of the Breit - Fermi Hamiltonian \cite{bbr}. 

It is also important to remember that a quark bound state is described as an effective degree of freedom ``smeared out'' by gluons and quark - antiquark pairs in such a way that some authors propose a density function of quarks in configuration space describing such potential delocalization effect \cite{god, ban, bra}.
The delocalization effect, in a simplified description like this one, is already achieved with the parameter $\mbox{w}$  of the equation (\ref{eq15}).

A finite width effect parameterization could be included.
It generates a mathematical mixing with the delocalization effect.
For simplicity, we develop the calculations just with one parameter ($\mbox{w}$) describing both effects mixed. 
        
\section{\label{sec4}Methods and calculations}

Considering that our prescription involves the study of a perturbative Hamiltonian that comprises all relativistic corrections appearing in Gromes prescription \cite{gro, luc}, we use simple calculations in opposition to more extensive methods like orthogonal collocation \cite{ld1} or some not so exact variational method.
We find the parameters choosing a simplification adopting the mean square minimization of experimental masses of eight resonances (four charmonia and four bottomonia) that are excellent input data \cite{rpp}.

Adopting the wave function $\psi(\bi{r})=R_{nl}(r)Y_{lm}(\hat{r})$, the radial equation for it is written like (\ref{eq17}):
\begin{equation} 
(T^{kin}+U_s(r)+U_v(r)-E_{nl})R_{nl}(r)=0. 
\label{eq17} 
\end{equation} 
 
The terms $U_s(r)$ and $U_v(r)$ involve all relativistic corrections following the Gromes prescription \cite{gro, luc}.
It permits us to obtain the eigenvalues $E_{nl}$ for all quantum numbers of a quark - antiquark system.

The proper choice of the radial functions depends, generically, on the problem to be solved.
Analyzing some propositions (3D harmonic oscillator and Coulomb wave functions) we employed the functions proposed by Fulcher \cite{ful}; these permit a simpler numerical treatment with less error possibilities when a numerical integration is eventually truncated \footnote{It provides a relative error of magnitude less than $10^{-20}$ when truncating the numerical integration.}.  

That function described in expression (\ref{eq18}) was also employed by other authors \cite{ld1, ful, ld2, ld3} and it presents a similar analytical behavior to the Schr{\"{o}}dinger wave function of the Coulomb potential:       
\begin{equation} 
R_{nl}(r)=\sqrt{\frac{(2 \beta)^3 k!}{(k+2l+2)!}}(2 \beta r)^l \,e^{(-\beta r)} {L_k}^{2l+2}(2 \beta r), 
\label{eq18} 
\end{equation}   
where ${L_k}^{2l+2}(\beta r)$ are the associated Laguerre polynomials and the principal quantum number $n$  of the bound state $n^{2s+1}l_J$ is written as $n\,=\,k \,+1$.

We are interested in obtaining a potential that describes, in a unique way, all the results of the interacting dynamics for charm systems as well as bottom systems, both with the same set of parameters.

We introduce two more input parameters related to the electronic transition rates through the equation (\ref{eq19}) \cite{luc}:   
\begin{equation} 
\frac{\Gamma _{e^-e^+}(\Upsilon)}{\Gamma _{e^-e^+}(\Psi)}=\frac{1}{4}\frac{{m_c}^2}{{m_b}^2}{\left(\frac{\beta _b}{\beta _c}\right)}^3 .
\label{eq19} 
\end{equation}  
 
The difference between charmonium and bottomonium resonances will depend on the parameters $\beta _c$  and $\beta _b$ in the wave functions.
Experimentally we have \cite{rpp}:
\begin{eqnarray} 
\Gamma _{e^-e^+}(\Psi)=5.26 \pm 0.37 \, \mbox{keV},\nonumber\\
\Gamma _{e^-e^+}(\Upsilon)=1.32 \pm 0.05 \, \mbox{keV},
\label{rate}
\end{eqnarray} 
indicating: 
\begin{equation}
\beta _c \approx \beta _b \left( \frac{m_c}{m_b}\right)^{\frac{2}{3}}.
\label{eq20}
\end{equation}

Resuming the treatment, the calculations developed here are parameterized for: the charm and bottom quark effective masses ($m_c$ and $m_b$), the wave function parameter $\beta _b$ and the potential parameters $\alpha$, $a$, $b$ and w \footnote{The w of the exponential expression that parameterizes the interaction vector term may impose a Coulomb behavior to this term of potential, depending on the energy levels that we are computing.}.
They are obtained through the above quoted mean square minimization.

\section{\label{sec5}Results for mass spectrum of charmonia and bottomonia}
  
In accordance with the systematics described before, using the wave function proposed \cite{ful}, we calculate the energy of $\eta_c$, $\Psi_1$, $\Psi_2$ and $\Psi_3$ charmonium states and of $\Upsilon_1$, $\Upsilon_2$, $\Upsilon_3$ and $\Upsilon_4$ bottomonium states, all with quantum numbers and experimental values (input data) described in tables \ref{tab2} and \ref{tab3}.
We also use the values of expression (\ref{rate}) as input data.
\begin{table}
\caption{\label{tab1}Phenomenological parameters for the charmonium - bottomonium unified effective model.} 
\begin{indented}
\item[]\begin{tabular}{@{}lllcclll} 
\br
$m_c$ & $=$ & $ 1.108 \, \mbox{GeV}$      & & & $\alpha _s$ & $=$ & $ 0.739 $  \\ 
$m_b$ & $=$ & $ 4.727 \, \mbox{GeV}$      & & & $a$ & $=$ & $ 0.0475\,\mbox{GeV}^2$  \\ 
$\beta _c$ & $=$ & $ 0.148 \, \mbox{GeV}$ & & & $b$ & $=$ & $ 0.088~\,\mbox{GeV}$  \\ 
$\beta _b$ & $=$ & $ 0.390 \, \mbox{GeV}$ & & & $\mbox{w}$ & $=$ & $ 0.550~\,\mbox{GeV}$ \\ 
\br
\end{tabular}
\end{indented}
\end{table}
\begin{table}
\caption{\label{tab2} Charmonium -  $\Psi$ mass spectrum calculated in the use of a Coulomb plus Linear Potential.\label{tab2}} 
\begin{indented}
\item[]\begin{tabular}{@{}lcrrr} 
\br
& & \multicolumn{3}{c}{Effective Mass (MeV)}\\
\cline{3-5}  
meson & $n^{2s+1}l_J$ &  Experiment$^{\rm a}$ & Theory & Difference\\ 
\mr
$\Psi_1$ & $1^3 S_0$ & 3096.9 &	3096 & $  -1$\\  
$\Psi_2$ & $2^3 S_0$ & 3686.0 & 3476 & $-210$\\ 
$\Psi_3$ & $3^3 S_1$ & 3769.9 & 3851 & $ +81$\\ 
$\Psi_4$ & $4^3 S_1$ & 4040 $~ \,$         & 4223 & $+183$\\ 
$\Psi_{(?)}$ & $5(?)^3 S_1$ & 4159 $~ \,$  & 4593 & $+434$\\ 
$\Psi_{(?)}$ & $6(?)^3 S_1$ & 4415 $~ \,$  & 4960 & $+545$\\ 
$\eta_c$ & $1^1 S_1$ & 2979.8 & 3093 & $+113$\\ 
$\chi_0$ & $1^3 P_0$ & 3417.3 & 3468 & $ +51$\\ 
$\chi_1$ & $1^3 P_1$ & 3510.5 & 3468 & $ -42$\\ 
$\chi_2$ & $1^3 P_2$ & 3556.2 & 3467 & $ -89$\\ 
?$^{\rm b}$ & $1^1P_1$ & ? & 3467 & ---\\ 
? & $2^1 P_1$ & ? & 3815 & ---\\ 
? & $2^3 P_0$ & ? & 3814 & ---\\ 
? & $2^3 P_1$ & ? & 3815 & ---\\ 
? & $2^3 P_2$ & ? & 3815 & ---\\ 
? & $3^3 P_0$ & ? & 4160 & ---\\ 
? & $3^3 P_1$ & ? & 4162 & ---\\ 
? & $3^3 P_2$ & ? & 4163 & ---\\ 
? & $1^1 D_2$ & ? & 3806 & ---\\ 
? & $2^1 D_2$ & ? & 4143 & ---\\ 
? & $1^3 D_1$ & ? & 3808 & ---\\ 
? & $1^3 D_2$ & ? & 3807 & ---\\ 
? & $1^3 D_3$ & ? & 3805 & ---\\ 
? & $2^3 D_1$ & ? & 4145 & ---\\ 
? & $2^3 D_2$ & ? & 4145 & ---\\ 
? & $2^3 D_3$ & ? & 4143 & ---\\ 
\br
\end{tabular}
\item[] $^{\rm a}$ Measurement errors with last significant digit order \cite{rpp}.
\item[] $^{\rm b}$ ? - Experimental data not accurately available.
\end{indented} 
\end{table}
\begin{table}
\caption{\label{tab3}Bottomonium - $\Upsilon$ mass spectrum calculated in the use of a Coulomb plus Linear Potential.\label{tab3}}  
\begin{indented}
\item[]\begin{tabular}{@{}lcrrr} 
\br
& & \multicolumn{3}{c}{Effective Mass (MeV)}\\
\cline{3-5}  
meson & $n^{2s+1}l_J$ &  Experiment & Theory & Difference\\ 
\mr
$\Upsilon_1$ & $1^3 S_1$ &  9460.4   &  9608 & $+149$\\  
$\Upsilon_2$ & $2^3 S_1$ & 10023.3   &  9931 & $ -92$\\ 
$\Upsilon_3$ & $3^3 S_1$ & 10353.3   & 10237 & $-116$\\ 
$\Upsilon_4$ & $4^3 S_1$ & 10580 $~ \,$        & 10533 & $ -47$\\ 
$\Upsilon_5$ & $5^3 S_1$ & 10865 $~ \,$        & 10825 & $ -40$\\ 
$\Upsilon_{(?)}$ & $6(?)^3 S_1$ & 11019 $~ \,$ & 11113 & $ +96$\\ 
$\chi_{b0} $ & $1^3 P_0$ &  9859.8 &  9811 & $ -49$\\ 
$\chi_{b1} $ & $1^3 P_1$ &  9891.9 &  9812 & $ -80$\\ 
$\chi_{b2} $ & $1^3 P_2$ &  9913.2 &  9812 & $-101$\\ 
$\chi'_{b0}$ & $2^3 P_0$ & 10231.1 & 10042 & $-189$\\ 
$\chi'_{b1}$ & $2^3 P_1$ & 10255.2 & 10043 & $-212$\\ 
$\chi'_{b2}$ & $2^3 P_2$ & 10268.5 & 10044 & $-224$\\ 
? & $1^1 S_0$ & ? &  9607 & --- \\ 
? & $1^1 P_1$ & ? &  9812 & --- \\ 
? & $2^1 P_1$ & ? & 10043 & --- \\ 
? & $3^3 P_0$ & ? & 10270 & --- \\ 
? & $3^3 P_1$ & ? & 10271 & --- \\ 
? & $3^3 P_2$ & ? & 10272 & --- \\ 
? & $1^1 D_2$ & ? &  9980 & --- \\ 
? & $2^1 D_2$ & ? & 10174 & --- \\ 
? & $1^3 D_1$ & ? &  9980 & --- \\ 
? & $1^3 D_2$ & ? &  9980 & --- \\ 
? & $1^3 D_3$ & ? &  9980 & --- \\ 
? & $2^3 D_1$ & ? & 10174 & --- \\ 
? & $2^3 D_2$ & ? & 10174 & --- \\ 
? & $2^3 D_3$ & ? & 10175 & --- \\ 
\br
\end{tabular} 
\end{indented}
\end{table} 

With the expressions found we calculate the phenomenological parameters described in table \ref{tab1} minimizing the function $\chi^2$, since we are interested in the parameter $\chi' \, = \,\sqrt{\frac{\chi^2}{n}}$ ($n$ is the number of input parameters).       
\begin{equation} 
\chi ^2= \sum_{i=1}^8{\left[\frac{({M_i}^{theo}-{M_i}^{exp})}{\Delta M_i }\right] }^2   
\end{equation} 

The term $\Delta M_i$ is the experimental error in the measurement of the mass ${M_i}^{exp}$. 
If $\Delta M_i$ is less than $3 \, MeV$, $\Delta M_i = 3 \, MeV$ is adopted minimizing the influence of any input data \cite{bra}. 
  
We found the minimum value $\chi' \, = \, 42.6$.  
 
In this way we obtain the parameters that characterize the quark - antiquark interactions in the charmonium - $\Psi$ and bottomonium - $\Upsilon$ resonances, obeying a Coulomb plus Linear behavior (see equation (\ref{eq16})). 
 
With those parameters we calculate the charmonium and bottomonium mass spectrum for several quantum numbers.
Some authors \cite{luc,qui,ld2} suggest that the calculation of transition rates should be done in parallel to mass spectroscopy but the results, confirmed by ours, corroborate the fact that the mass spectroscopy is not the best way to do so.
The mass spectroscopy results are described in tables \ref{tab2} and \ref{tab3}.
 
With these values, considering just the data with quantum numbers exactly confirmed through the experimental measurements, we observe a mean deviation of $3.54 \, \%$ in the charmonium mass spectrum and a mean deviation of $1.14 \, \%$ in the bottomonium mass spectrum, when we compare with the experimental data.

Since we were working all the time with a reduced and unified number of parameters for charm and bottom families, the results obtained describe the mass spectroscopy in a way as good as other results \cite{god, ban, bra, sem} obtained with a more complex and extensive analysis.

Other aspects also deserve some emphasis.
It is clear, with corroboration of some results already obtained, that the spectroscopy study with different sets of parameters for each meson family would provide better results despite the fact that experimental data with rigorously accurate quantum numbers are not sufficient to establish a trusty statistics.

Attending to the accuracy of the nonrelativistic approximation adopted for the kinetic term, it is quite natural that we observe a minor mean deviation for bottomonium states since charmonium states could be considered semirelativistic.

Another important observation is that, like results of other authors \cite{luc, god, ban, bra, sem, qui}, there is an experimental disagreement relative to degenerancy in total angular momentum $J$.
Observing the results in tables \ref{tab2} and \ref{tab3}, we can conclude that Breit - Fermi interaction terms are not sufficiently accurate to describe states with different values of $J$ $(J \, = l-1,\, l \,\, \mbox{and} \,\, l+1)$.

\section{\label{sec6}Conclusion}

When a study of mass spectroscopy of quark bound states is developed, the first aspect that will establish its success is obviously the correspondence of the theoretical model to the experimental measurements.
In this way, the potential model developed and applied here permits a broad and accurate description of the mass spectrum of charmonium and bottomonium systems when we attempt to the mean deviation of all resonances.

Adopting an effective Hamiltonian formalism built on a restrict set of eight parameters\footnote{We remind that $\beta_c$ and $\beta_b$ are linked by expression (\ref{eq20}), so there are only seven independent parameters.} proved to be a significant simplification in the calculations here developed.
Other authors \cite{god, ban, sem, bra} employ more complex developments that often use more than twenty parameters.
In spite of the use of some complex developments, none of them is capable to describe states with different values of $J$ $(J \, = l-1,\, l \,\, \mbox{and} \,\, l+1)$ accurately.
Some of them \cite{ban, ld1, ld2} do not even bother about such calculations involving the total angular momentum quantum number.
Others \cite{god, sem, bra} obtain results with deviations as large as those presented in this paper.  

The first simplification occurs because some nonlocal terms that would appear in the explicit potential could be ``absorved'' by the linear parameters.
The mass spectroscopy of hadrons following a potential model is not the best way to indicate the exact form of the nonlocal terms.

Searching for a simpler model, we succeed in describing the kinetic term in a nonrelativistic approximation  in which, differently to other models \cite{god, ban, ld1, ful, ld2}, we adopt a kinetic term accurate in the energy region below $10 \, \mbox{GeV}$.
It minimizes the difference to the exact kinetic term.

The advantage of such procedure is that the obtained values for ${\bi{p}_0}^2$ are independent of any energy state used as input data.
In this way the extension of using the same value of ${\bi{p}_0}^2$ for any other resonances of that particular bound state is immediate.
In comparison, Jaczko and Durand do not extend their calculation in \cite{ld2} beyond the four lower energy states which otherwise were extensively studied in their paper.
The quoted authors \cite{ld2} had chosen to keep the kinetic term depending on one more parameter ($\epsilon^{\prime}$) in the effective Hamiltonian.

All that simplification together with a reduced number of eight parameters allows for the use of an also reduced number of input data.
With this, the higher energy levels do not present a direct constraint with the experimental data, a fact that would force a better agreement.

It should be emphasized that the set of parameters is common to both families of charmonia and bottomonia.
Lower deviations from the experimental values could have been achieved if a different set of parameters had been adopted for each family separately. 

Despite the relative success, some questions concerning potential models yet remain unanswered: Why do we not try to solve directly (numerically) a relativistic bound state equation instead of following an effective model?

Generically, we can conclude that there is a great number of possibilities to solve a quark bound state problem, sometimes through analytically or numerically complicated calculations, but none of them is exact.
We hope that some direct or indirect results of QCD can provide new guides to produce more accurate phenomenological models.          
 
\ack

The authors wish to thank CAPES and FAPESP for financial support.
 

\end{document}